\documentclass[conference]{IEEEtran}
\IEEEoverridecommandlockouts
\usepackage{cite}
\usepackage{amsmath,amssymb,amsfonts}
\usepackage{algorithmic}
\usepackage{algorithm}
\usepackage{graphicx}
\usepackage{textcomp}
\usepackage{xcolor}
\usepackage{makecell}
\usepackage{multirow}

\usepackage{hyperref}
\usepackage{booktabs}

\newcommand{\upp}[1]{\color[rgb]{0, 0.653, 0}#1}
\newcommand{\dow}[1]{\color[rgb]{0.753,0,0}#1}

\newcommand{\inc}[1]{\upp\mathbf{\blacktriangle #1\%}}
\newcommand{\dec}[1]{\dow\mathbf{\blacktriangledown #1\%}}

\def\BibTeX{{\rm B\kern-.05em{\sc i\kern-.025em b}\kern-.08em
    T\kern-.1667em\lower.7ex\hbox{E}\kern-.125emX}}
\begin{document}

\title{DrLLM: Prompt-Enhanced Distributed Denial-of-Service Resistance Method with Large Language Models\\
\thanks{$^{\ast}$Corresponding Author}
\thanks{This work is supported in part by Special Project for Industrial Foundation Reconstruction and High Quality Development of Manufacturing Industry(No. TC230A076-13).}
}

\author{\IEEEauthorblockN{1\textsuperscript{st} Zhenyu Yin$^{1,3,\ast}$, 2\textsuperscript{nd} Shang Liu$^{1,2,3}$, 3\textsuperscript{nd} Guangyuan Xu$^{1,2,3}$}
\IEEEauthorblockA{
$^1$\textit{Shenyang Institute of Computing Technology, Chinese Academy of Sciences} \\
$^2$\textit{University of Chinese Academy of Sciences} \\
$^3$\textit{Liaoning Key Laboratory of Domestic Industrial Control Platform Technology on Basic Hardware and Software} \\
congmy@163.com, \{liushang221, xuguangyuan18\}@mails.ucas.ac.cn}}

\maketitle

\begin{abstract}
The increasing number of Distributed Denial of Service (DDoS) attacks poses a major threat to the Internet, highlighting the importance of DDoS mitigation. Most existing approaches require complex training methods to learn data features, which increases the complexity and generality of the application. In this paper, we propose DrLLM, which aims to mine anomalous traffic information in zero-shot scenarios through Large Language Models (LLMs). To bridge the gap between DrLLM and existing approaches, we embed the global and local information of the traffic data into the reasoning paradigm and design three modules, namely Knowledge Embedding, Token Embedding, and Progressive Role Reasoning, for data representation and reasoning. In addition we explore the generalization of prompt engineering in the cybersecurity domain to improve the classification capability of DrLLM. Our ablation experiments demonstrate the applicability of DrLLM in zero-shot scenarios and further demonstrate the potential of LLMs in the network domains. The implementation of DrLLM is available at \url{https://github.com/liuup/DrLLM}.
\end{abstract}


\section{Introduction}
The Distributed Deniel-of-Service (DDoS) threats in cyber space are becoming increasingly sophisticated and complex, posing a threat to critical Internet devices and systems including routers, switches or firwalls. Today, the scale of the Internet and the Internet of Things (IoT) is much larger than before, and the demand for network security measures is also increasing\cite{ferrag2024generative}. Although the Instrusion Detection System (IDS) countermeasures against DDoS are developing rapidly, network attack events are still increasing. Imperva released their report\cite{impervareport}, claiming the DNS attacks increased in number by 215\% when comparing H1 2024 with the same period in 2023 and the total number of recorded DDoS attacks surged 111\% compared to the previous year. 

Over the past few years, we have witnessed Large Language Models (LLMs), such as ChatGPT\cite{achiam2023gpt}, Llama\cite{dubey2024llama}, Deepseek\cite{bi2024deepseek} and the other models\cite{jiang2023mistral, reid2024gemini} have make siginificant influence in artificial intelligence (AI), these LLMs include leading and fancy model architecture, and were trained from mega-billion tokens which compressed with domain knowledge. As generative models, it can not only generate a variety of creative responses to meet user needs, but it is also easy to use and only requires a simple prompt to drive it\cite{kojima2022large}. The fantastic capability has been proved on neural language processing (NLP), multimodality fusion and other fields\cite{yin2023survey}\cite{wang2024cog}, which even laying the foundation stone for Artifical General Intelligence (AGI). 

Inspired by the success of LLMs in many fields, especially textual information mining. In this paper, we introduce DDoS Resistance Large Language Model (DrLLM) which contains three modules, a novel method that can efficiently extract the vast network traffic knowledge contained in pre-trained LLMs via progressive Role Reasoning pipeline by prompting. With our prompting strategy, by aligning structed network streaming data into text space to preserve the full semantics, we can embed global information and local information in prompt template to enhance the classifiction capability. We evaluated the performance of DrLLM with the public dataset CICDDoS2019\cite{sharafaldin2019developing}, and we also conducted ablation experiments on zero-shot scenarios to validate the functionality of DrLLM's core modules.

We present the following main research contributions:
\begin{itemize}
    \item We propose a progressive DDoS traffic detection framework based on LLMs which is called DrLLM, which can efficiently classify network flow data and improve the interpretability of the classification basis.
    \item We innovatively use Knowledge Embedding to embed the global information of the data into prompting template, and conduct ablation experiments to demonstrate its effectiveness.
    \item We carefully crafted multiple prompting templates and Constrain-of-Deviation (CoD) and Chain-of-Thought (CoT) has been used in Token Embedding module to embed the local information and enhance the classification performance of LLMs.
\end{itemize}

\section{Background and Related Works}

\subsubsection{Cyber IDS and Method} For DDoS traffic detection and classification tasks, existing IDSs are usually based on neural network architectures and the final results reached state-of-the-art (SOTA). Yisroel Mirsky et al.\cite{mirsky2018kitsune} presented Kitsune based on autoencoders to detect abnormal traffic data. Lucid\cite{doriguzzi2020lucid} was a lightweight deep learning solution  in recource-constrained environments. IDS-INT\cite{ullah2024ids} used transformer-based transfer learning for imbalanced network traffic and detect minority attacks. In practical scenarios, they are mostly deployed on the data plane or control plane of a programmable switch\cite{barradas2021flowlens, liu2021jaqen} to handle large-scale hybrid and dynamic traffic attacks, and differentiate between normal and abnormal patterns.

\subsubsection{LLMs Enhanced Cybersecurity} Large Language Models (LLMs) can excel across various domains, including Neural Language Processing (NLP) and Computer Vision (CV)\cite{achiam2023gpt,dubey2024llama}, and recent research in time series forecasting\cite{jin2024timellm, gruver2023large}, information mining\cite{manvi2024geollm, wu2024exploring} has demonstrated the effectiveness of pre-trained models that can be fine-tuned for various multimodality tasks\cite{liu2023pre}. In addition, some methods have been proposed and applied to optimize LLMs\cite{kojima2022large, liu2021generated}. Current research demonstrates the potential of leveraging LLMs in network security, it is not only used to detect software vulnerabilities\cite{ferrag2024securefalcon}, but also to detect DDoS traffic\cite{li2024dollm} and generate mitigation strategies\cite{wang2024shieldgpt} to safeguard against evolving cyber threats\cite{zhang2024llms}.

\section{Overview and Design}

\begin{figure*}
\centering
\includegraphics[width=\textwidth]{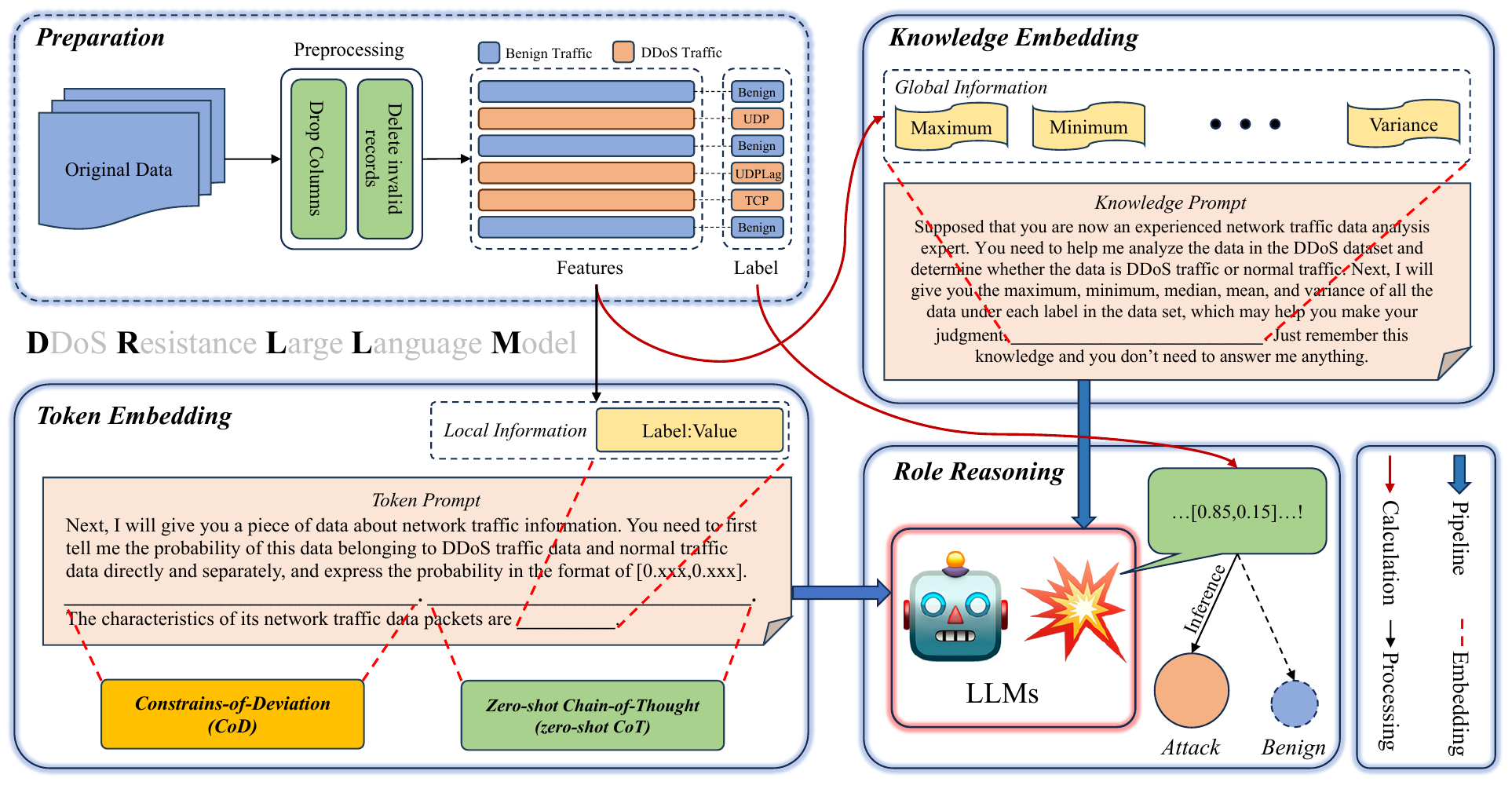}
\caption{The overall framework of DrLLM, we have divided the framework into four main modules: preparation module, knowledge embedding module, token embedding module and role reasoning module.} \label{overall}
\end{figure*}

DrLLM has two primary objectives: 1. Preserve the same semantic information in tabular data and text data, and establish a new paradigm to align tabular data to text modality. 2. Through specific feature engineering, relying on LLMs to mine network flow data, build progressive reasoning pipeline. In order to achieve these two goals, we designed three main modules for DrLLM: Knowledge Embedding, Token Embedding, and Role Reasoning. The overview of DrLLM architecture is shown in Fig~\ref{overall}.

% The Knowledge Embedding module is primarily tasked with the extraction of the global information of the dataset and their subsequent embedding into the Knowledge Prompt. The objective is to augment the LLMs' expert knowledge, mitigate output hallucinations, and enhance classification accuracy. The Token Embedding module is primarily tasked with encoding network flow data, aligning tabular data to the text modality, and leveraging multiple technologies to enhance accuracy. The Role Reasoning module is responsible for integrating the aforementioned two modules, constructing the reasoning pipeline, and obtaining classification results through LLMs' output. In general, our objective is to facilitate the analysis of network traffic data by LLMs through the construction of associations between global prior knowledge and local information, thereby enabling the accurate classification of data. 

\subsection{Knowledge Embedding}
We assign expert assumptions at the beginning of Knowledge Prompt to converge the reasoning ability of large models to the network field and clarify specific network traffic flow classification tasks. In the zero-shot scenario, for LLMs, the representation of a single data point alone may lack prior information for judgment, because this ignores the global data feature distribution of the dataset. After the data preprocessing, we calculate the 5-tuple characteristics $G_{m}$ (maximum, minimum, median, mean and variance) for each column in dataset as the global information. This simple and effective method can serve as prior knowledge for LLMs and inform the global data distribution of the dataset in advance. To reduce the output length in this module and increase the efficiency of reasoning, we constrain its output at the end position of the prompt to reduce the output content of this step.

The detailed process of Knowledge Embedding is as follows: Assume that for the dataset $D_{n\times m}(X)=[{X_{1},X_{2},...,X_{n}}]^\top$, where $X_{n}=(x_{1},x_{2},...,x_{m})$ and features are $Features=(F_{1},F_{2},,...F_{m})$, for every column $X_{m}$, to calculate the maximum $a$, minimum $b$, median $c$, mean $d$ and variance $v$, and finally get $Global_{m}=[G_{1},G_{2},...,G_{m}]^\top$, where $G_{j}=(a_{j},b_{j},c_{j},d_{j},v_{j})_{\{j=1,2,...,m\}}$. Then construct the string:

\begin{equation}
K(G)=Concat(Global_{m}\oplus S)\label{eq}
\end{equation}
where $Concat$ means converting the elements in the vector into strings and concatenating them with $S=(``Max",``Min",``Median",``Mean",``Variance")^\top$.
Finally we embed $K$ into Knowledge Prompt ($KP$).

\subsection{Token Embedding}
In order to enable LLMs such as ChatGPT to understand network flow data, we designed a module called Token Embedding, which uses text to represent the specified information flow. We choose LLMs as our inference backbone, and we will introduce the design of the Token Embedding module below and describe in detail how Constrain-of-Deviation (CoD) and Zero-shot Chain-of-Thought(CoT) in Prompt Engineering improve and enhance the information extraction and classification accuracy of LLMs.

\subsubsection{Constrain-of-Deviation (CoD)} LLMs with decoder-only architecture are very suitable for generative tasks. They can flexibly generate text and code information, but their current structured content (such as XML or JSON) output capabilities are still weak\cite{liu2024we}. When LLMs are integrated into our inference workflows, we need to contrain the outputs to follow specific standards. For the binary classification task of network traffic flow, in order to ensure consistency and predictability of output and reduce the hallucination, we created CoD Prompt to constrain the model to output according to instructions. This simple but very effective method allows us to use regular expressions to extract classification results from the output of LLMs. 

\subsubsection{Zero-shot Chain-of-Thought(CoT)} Zero-shot-CoT\cite{wei2022chain}is a zero-shot template-based prompting for chain of thought reasoning. Different from few-shot prompting with examples, it can enlighten the LLMs to make progressive thinking and perform complex reasoning for each instruction in a simple but effective pattern. We use CoT to guide the LLMs to progressively analyze the various characteristic values of network flow data and help improve calculation accuracy. 

\subsubsection{Token Prompt} In the Token Prompt ($TP$), the specific network traffic task representation is initially indicated and the LLMs are directed to provide a probabilistic judgment based on each piece of data. We then construct Token Prompt based on CoD and Zero-shot-CoT, and use it as the benchmark input of the Token Embedding module. For tabular data of network flows, we treated as a new data modality and embed it into prompt to align it to the text modality, and then feed it into LLMs for reasoning. The specific process is as follows: Given a dataset $D_{n \times m}(X)=[X_{1},X_{2},...,X_{n}]^\top$, which owns $Feature_{m}=[F_{1},F_{2},...,F_{m}]$, for every $X_{n}=[x_{1},x_{2},...,x_{m}]$, we construct the text string $TP_{i}(F_{j},X_{i})=Concat(F_{j}:X_{i})_{\{j=1,2,...,m,i=1,2,...,n\}}$ for each piece of data, and then embed $TP_{i}$ into $TP$. 

\subsection{Role Reasoning} \label{rolereasoning}
Based on Knowledge Embedding and Token Embedding, we build the progressively Role Reasoning pipeline on top of these two modules. In the first step, we embed the global information into the Knowledge Prompt $KP$, and feed $KP$ as the first step input into the $LLM$ to obtain the output $R^{1}_{i}$. In the second step, $R^{1}_{i}$ and Token Prompt $TP_{i}$ are sequentially fed into the $LLM$ for reasoning to obtain the output $R^{2}_{i}$. Finally, we use regular expressions to obtain the classification probability $P_{i}$ from $R^{2}_{i}$ and calculate the evaluation result $E_{i}$ with the true value $\hat P_{i}$ of the data $X_{i}$. 

% \begin{algorithm}
%     \caption{Role Reasoning Pseudocode}
%     \label{rolereasoning}
%     \begin{algorithmic}[1]
%         \REQUIRE{backbone $LLM$, dataset $D_{n\times m}$}
%         \ENSURE{evaluation $E$ of dataset $D_{n\times m}$}
        
%         \STATE $KP \gets preset$;
%         \STATE $S \gets (``Max",``Min",``Median",``Mean",``Variance")^\top$;
%         \FORALL {column $X_{j}$ in dataset $D_{n\times m}$}

%             \STATE $G_{j} \gets \{a_{j}=max(X_{j}),b_{j}=min(X_{j}),c_{j}=median(X_{j}),d_{j}=mean(X_{j}),v_{j}=variance(X_{j})\}_{\{j=1,2,...,m\}}$;
%             \STATE $KP$ += $concat(G_{j}, S)$;
        
%         \ENDFOR
        
%         \FORALL {data $X_{i}$ in dataset $D_{n\times m}$}
%             \STATE $TP_{i}\gets preset$;
        
%                 \FORALL {feature $F_{j}$ in features $D_{n\times m}$}
%                 \STATE $TP_{i}$ += concat($F_{j}$,$X_{i}$);
%                 \ENDFOR
        
%             \STATE $R^{1}_{i} \gets LLM(K)$;
%             \STATE $R^{2}_{i} \gets\ LLM(R^{1}_{i}, TP_{i})$ \textcolor{blue}{\COMMENT{Inference in order}}
            
%             \STATE $\hat P_{i} \gets REGEX(R^{2}_{i})$;
%             \STATE $E_{i} \gets eval(P_{i},\hat P_{i})$;
        
%         \ENDFOR
%         \STATE return $E=(E_{1},E_{2},...,E_{n})$;
%     \end{algorithmic}
% \end{algorithm}

\section{Experimental Evaluation}
\subsection{Datasets and Data Preprocessing}
The DrLLM prototype is implemented with Python 3.11.9. We conducted extensive experiments for DrLLM on CICDDoS2019 dataset\cite{sharafaldin2019developing}. The original dataset contains approximately 29GB of DDoS traffic data and benign traffic data, each with 88 features. According to the experimental needs, we preprocessed the data and deleted each row of data containing ``NaN" or ``Inf" values and we converted all DDoS traffic label to "Attack" and make benign traffic data unchanged. 
\newcommand{\bs}[1]{\color[rgb]{0.117, 0.447, 0.999}#1}
\newcommand{\ws}[1]{\color[rgb]{0.753,0,0}#1}

\begin{table*}[h]
	\centering

	\caption{Accuracy of DrLLM with Different LLM Backbones}
	\label{tab1}
	\renewcommand{\arraystretch}{1.2}
	\begin{tabular}{c|c|ccccc}
                % \cline{1-7}
                \toprule[1pt]

                \textbf{Backbones} & \textbf{Metrics$^{1}$} & \makebox[0.1\textwidth][c]{$P_{0}$} & \makebox[0.1\textwidth][c]{$P_{1}$} & \makebox[0.1\textwidth][c]{$P_{2}$} & \makebox[0.1\textwidth][c]{$P_{3}^{'}$} & \makebox[0.1\textwidth][c]{$P_{3}$} \\
                \cline{1-7}
                
                % \multirow{3}{*}{GPT$^{1}$}
                \multirow{3}{*}{GPT-4o-mini}
			& F1 & $0.6573_{\dec{14.99}}$ & $0.7024_{\dec{9.16}}$ & $0.7197_{\dec{6.92}}$ & $0.7510_{\dec{2.87}}$ & \bs $0.7732$ \\
			& Recall & $0.9033_{\inc{19.31}}$ & $0.6219_{\dec{17.86}}$ & $0.6620_{\dec{12.56}}$ & $0.7308_{\dec{3.47}}$ & \bs $0.7571$ \\
			& AUC & $0.7208_{\dec{13.62}}$ & $0.7797_{\dec{6.57}}$ & $0.7957_{\dec{4.65}}$ & $0.8254_{\dec{1.09}}$ & \bs $0.8345$ \\
			\cline{1-7}

                % \multirow{3}{*}{Llama3$^{2}$}
                \multirow{3}{*}{Llama3-70b}
			& F1 & $0.5117_{\dec{13.33}}$ & $0.5275_{\dec{10.65}}$ & $0.2951_{\dec{50.02}}$ & $0.5403_{\dec{8.49}}$ & \bs $0.5904$ \\
			& Recall & $0.3567_{\dec{19.21}}$ & $0.3744_{\dec{15.20}}$ & $0.1737_{\dec{60.66}}$ & $0.5037_{\inc{14.09}}$ & \bs $0.4415$ \\
			& AUC & $0.6613_{\dec{18.60}}$ & $0.8543_{\inc{5.15}}$ & $0.7614_{\dec{6.28}}$ & $0.6789_{\dec{16.43}}$ & \bs $0.8124$ \\
			\cline{1-7}

                % \multirow{3}{*}{Qwen2$^{4}$}
                \multirow{3}{*}{Qwen2-57b-a14b-instruct}
                & F1 & $0.6525_{\dec{12.25}}$ & $0.6396_{\dec{13.98}}$ & $0.6968_{\dec{6.29}}$ & $0.7090_{\dec{4.65}}$ & \bs  $0.7436$ \\
			& Recall & $0.8307_{\inc{9.95}}$ & $0.7448_{\dec{1.41}}$ & $0.7383_{\dec{2.27}}$ & $0.7554_{\dec{0.01}}$ & \bs $0.7555$ \\
			& AUC & $0.6984_{\dec{13.69}}$ & $0.7038_{\dec{13.02}}$ & $0.7584_{\dec{6.27}}$ & $0.7310_{\dec{9.66}}$ & \bs $0.8092$ \\
                \cline{1-7}

                % \multirow{3}{*}{Deepseek$^{3}$}
                \multirow{3}{*}{Deepseek-chat-v2}
                & F1 &  $0.7672_{\dec{9.73}}$ & $0.7904_{\dec{7.00}}$ & $0.8114_{\dec{4.53}}$ & $0.7759_{\dec{8.71}}$ & \bs $0.8499$ \\
			& Recall & $0.6507_{\dec{20.77}}$ & $0.6994_{\dec{14.84}}$ & $0.6882_{\dec{16.21}}$ & $0.7256_{\dec{11.65}}$ & \bs $0.8213$ \\
			& AUC & $0.8710_{\dec{5.45}}$ & $0.8853_{\dec{3.90}}$ & $0.8927_{\dec{3.09}}$ & $0.8787_{\dec{4.61}}$ & \bs $0.9212$ \\

                \bottomrule[1pt]
        % \multicolumn{6}{l}{$^{1}$Denotes GPT-4o-mini. $^{2}$Denotes Llama3-70b.} \\
        % \multicolumn{6}{l}{$^{3}$Denotes Deepseek-chat-v2. $^{4}$Denotes Qwen2-57b-a14b-instruct.} \\
        \multicolumn{7}{l}{$^{1}$We highlight the $P_{3}$ by {\bs{$\bullet$}}.} \\
        \multicolumn{7}{l}{$^{2}$If other prompts' perform better than $P_{3}$ on a metric we denote it by {\upp{$\bullet$}}, if not as well we denote it by {\dow{$\bullet$}}.}
        
	\end{tabular}
\end{table*}

\subsection{LLM Backbones}
We evaluate DrLLM in zero-shot scenarios by comparing it with four state-of-the-art LLMs, including GPT-4o-mini\cite{gpt4omini}, Llama3-70b\cite{dubey2024llama}, Deepseek-chat-v2\cite{bi2024deepseek} and Qwen2-57b-a14b-instruct\cite{yang2024qwen2}. We directly call their official application programming interfaces (APIs) in non-streaming mode to enable rapid response and kept the parameters as official default. 

\subsection{Prompts and Evaluation Metrics}
As mentioned in Section \ref{rolereasoning}, we use regular expressions to extract classification probabilities from the output and constrain them using CoD. However, during the evaluation process, we found that the LLMs occasionally outputs in an incorrect format. In order to verify the effectiveness of multiple modules of DrLLM, we carefully crafted four prompt templates $P_{0}=BP(Basic Prompt)$, $P_{1}=BP+CoD$, $P_{2}=BP+CoD+CoT$, $P_{3}^{'}=KP+BP+CoD$, $P_{3}=KP+BP+CoD+CoT$ for comparison. Due to space constraints, we will present the full prompt templates in the open source repository.

In addition, in order to evaluate the performance of DrLLM on multiple models and multiple prompts, in this paper, we adopted F1, Recall and AUC as the evaluation metrics during our experiments.

\subsection{Ablation Study}
In order to analyze the effectiveness of DrLLM, we conduct a series of ablation experiments with other state-of-the-art LLMs on our proposed modules. As shown is Table \ref{tab1}.

Overall, Deepseek-chat-v2 performs the best and outperforms almost all other models under both $P_{0}$ and $P_{3}$, with F1 reaching $0.8499$ and AUC $0.9212$ under $P_{3}$, and also performs stably under $P_{1}$ and $P_{2}$, which we also use as the backbone of DrLLM. While the worst LLM is Llama3-70b, which is almost unusable in zero-shot scenario. In addition, all the models almost increase in each scenarios after adding our designed $KP$, $CoD$ and $CoT$, and the classification ability is greatly enhanced in zero-shot scenarios, which also proves the effectiveness of Knowledge Embedding and Token Embedding in DrLLM from the side.

\subsection{Stability Analysis}
During the evaluation process, we checked the outputs of multiple models and found that the outputs of LLMs were not stable and occasionally exhibited unconstrained abnormal behaviors. We collected the abnormal outputs and defined the anomalies as follows:

\subsubsection{Confidence Bias} We denotes this problem as $L_{1}$. In the classification, the model gives confidence for both abnormal and normal traffic, but the sum of their probability results is not 1. This type of problem is more frequent in the case of $P_{0}$, but after adding $CoD$, the probability of such problems is greatly reduced, which also indirectly proves the effectiveness of $CoD$ in DrLLM in constraining the output of large models.

\subsubsection{Confidence Lost} We denotes this problem as $L_{2}$. LLMs mean that analysis is impossible and there is a lack of confidence in the predicted probabilities based on the data.

Table \ref{tab2} demonstrates the probability of occurrence of the $L_{1}$ and $L_{2}$ in each scenarios. In the $P_{0}$, the $L_{1}$ of GPT-4o-mini has the highest probability $0.8760$, in contrast to DeepSeek-chat-v2 which has the best stability of the $L_{1}$ at only $0.0309$. 
Overall, comparing to other LLMs, Deepseek-chat-v2 performs most stably in multiple scenarios, but after adding of multiple modules, the $L_{1}$ of GPT-4o-mini drops rapidly to $0.0014$ and the $L_{2}$ to $0.0153$, which is comparable to the performance of Deepseek-chat-v2, which is a side-effect of proving the DrLLM multiple modules' effectiveness.

\begin{table}[h]
	\centering

	\caption{Abnormal Output of backbones.}
	\label{tab2}
	\renewcommand{\arraystretch}{1.2}
	\begin{tabular}{c|c|ccccc}
                \toprule[1pt]

                \textbf{Backbones} & \textbf{Metrics$^{5,6}$} & $P_{0}$ & $P_{1}$ & $P_{2}$ & $P_{3}^{'}$ & $P_{3}$ \\
                
                \cline{1-7}

                \multirow{2}{*}{GPT$^{1}$} 
 			& $L_{1}$ & $87.60$ & $18.07$ & $0.37$ & $0.33$ & \bs$0.14$ \\
			& $L_{2}$ & $3.40$ & $3.02$ & $1.92$ & $1.84$ &\bs $1.53$ \\
			\cline{1-7}

                \multirow{2}{*}{Llama3$^{2}$}
			& $L_{1}$ & $28.29$ & \bs{$0$} & $0.03$ & $0.07$ &$0.10$ \\
			& $L_{2}$ & \bs{$0$} & $0.48$ & \bs{$0$} & $0.17$ &$0.10$ \\
			\cline{1-7}
  
                \multirow{2}{*}{Qwen2$^{3}$}
                & $L_{1}$ & $53.46$ & $1.29$ & $0.65$ & $0.30$ & \bs{$0.27$} \\
			& $L_{2}$ & $14.03$ & $9.31$ & $2.45$ & $0.97$ & \bs{$0.92$} \\
                \cline{1-7}

                \multirow{2}{*}{Deepseek$^{4}$}
                & $L_{1}$ & $3.09$ & \bs{$0$} & \bs{$0$} & $1.27$ &$0.20$ \\
			& $L_{2}$ & $7.07$ & $3.09$ & $1.46$ & $1.19$ & \bs{$0.90$} \\
                \bottomrule[1pt]
            \multicolumn{6}{l}{$^{1}$Denotes GPT-4o-mini. $^{2}$Denotes Llama3-70b.} \\
            \multicolumn{6}{l}{$^{3}$Qwen2-57b-a14b-instruct. $^{4}$Denotes Denotes Deepseek-chat-v2.} \\
            \multicolumn{6}{l}{$^{5}$We highlight the best value by {\bs{$\bullet$}}.} \\
            \multicolumn{6}{l}{$^{6}$The unit of $L_{1}$ and $L_{2}$ is percentage [\%]. Lower is better.} 
            
	\end{tabular}
\end{table}

\section{Conclusion and Future Work}
In this paper, we proposed a DDoS resistance method based with Large Language Models which is called DrLLM. We embed the global information and local information with Knowledge Embedding and Token Embedding modules. We design the Progressive Role Reasoning for classification and explore the application of DrLLM in zero-shot scenarios with extensive experiments. The ablation experiments have demonstrated the effectiveness of the individual modules in DrLLM and proved the universality of LLMs in the networking domain. In the future, we will explore the application of retrieval-augmented generation (RAG) in LLMs.

\section*{Acknowledgment}
The authors wish to thank the reviewers for their helpful comments and suggestions. This work is supported in part by Special Project for Industrial Foundation Reconstruction and High Quality Development of Manufacturing Industry(No. TC230A076-13).

\bibliographystyle{IEEEtran}
\bibliography{./refer.bib}

\end{document}